Growth and characterization of R(O,F)BiS$_2$ (R = La, Ce, Pr, Nd) superconducting single crystals

Masanori Nagao[a,b,*]

[a]*University of Yamanashi, 7-32 Miyamae, Kofu, Yamanashi 400-8511, Japan*

[b]*National Institute for Materials Science, 1-2-1 Sengen, Tsukuba, Ibaraki 305-0047, Japan*

*Author

Masanori Nagao

Postal address: University of Yamanashi, Center for Crystal Science and Technology

Miyamae 7-32, Kofu 400-8511, Japan

Telephone number: (+81)55-220-8610

Fax number: (+81)55-254-3035

E-mail address: mnagao@yamanashi.ac.jp

**Abstract**

R(O,F)BiS$_2$ (R: La, Ce, Pr, Nd) superconducting single crystals with different F concentrations are grown using a CsCl/KCl flux. The obtained 1-2 mm-sized single crystals have a plate-like shape and are cleavable along the *ab*-plane. The crystal structure is tetragonal of space group *P*4/*nmm* (#129). The as-grown single crystals exhibit superconductivity at around 3-5 K. The superconducting transition temperature increases with decreasing the ionic radius of the R element. The superconducting anisotropies of the R(O,F)BiS$_2$ single crystals are estimated to be 30-60 according to the effective mass model, whereas the anisotropies for Ce(O,F)BiS$_2$ single crystals are 13-21. The *c*-axis transport measurements of a single crystal of Pr(O,F)BiS$_2$ under a magnetic field parallel to the *ab*-plane reveal a "lock-in" state attributed to the Josephson vortex flow. Furthermore, the Ce(O,F)BiS$_2$ single crystals exhibit a magnetic order at the temperature range of 5-7 K that apparently coexist with superconductivity below approximately 3 K.

**Keywords**

# 1. Introduction

Representative examples of $BiS_2$-layered superconductors include $Bi_4O_4S_3$ [1], $R(O,F)BiS_2$ (R: La, Ce, Pr, Nd, Yb) [2-6], $AFBiS_2$ (A: Sr, Eu) [7-9], $La_{1-b}M_bOBiS_2$ (M: Ti, Zr, Hf, Th) [10], $Eu_3F_4Bi_2S_4$ [11], and $BiO_{1-c}F_cBiS_2$ [42,43]. Since $R(O,F)BiS_2$ shows similarity to the crystal structure of superconducting $RnO_{1-d}F_dFeAs$ (Rn: rare earth elements) [12,13], much effort has been focused on the growth of single crystals thereof [14-18]. Alkali chloride fluxes, such as LiCl and NaCl, are generally useful for the growth of single crystals. However, $R(O,F)BiS_2$ cannot be synthesized using LiCl and NaCl fluxes in a vacuum quartz tube at high temperatures because these alkali chlorides would corrode the quartz tube. A mixture of CsCl and KCl (CsCl/KCl) is a promising flux, as it exhibits little reactivity to quartz tubes at high temperatures [19-21].

Various compositions of these single crystals are highly desired for investigating the effects of different R elements and F concentrations on the intrinsic properties of $R(O,F)BiS_2$. When using polycrystalline samples, the intrinsic properties are masked by the impurities and grain boundaries and the anisotropic properties cannot be measured. Single crystals are also necessary for surface analyses by scanning tunnel microscopy (STM) and the fabrication of mono-layered devices [15,22,23], though these are not detailed in this paper.

This review focuses on the growth and characterization of R(O,F)BiS$_2$ (R = La, Ce, Pr, Nd) single crystals, which show not only superconductivity but also magnetism and high anisotropy. The growth of R(O,F)BiS$_2$ single crystals was performed using CsCl/KCl flux. The composition and superconducting properties were investigated to characterize the properties of the single crystals. The crystal structural analysis was performed by single-crystal X-ray diffraction (XRD), which revealed the relationship between the crystal structures and F concentrations of the crystals. The transport properties were examined to understand the intrinsic properties of these materials. The superconducting anisotropies of R(O,F)BiS$_2$ single crystals were estimated and showed high values. The high anisotropy in layered superconductors demonstrates their potential use as intrinsic Josephson junctions, considering the weak coupling among BiS$_2$-RO(F)-BiS$_2$ (superconducting-normal-superconducting) layers along the $c$-axis [24,25]. Consequently, the $c$-axis transport property of the Pr(O,F)BiS$_2$ single crystal was investigated. The coexistence of superconductivity and magnetism in Ce(O,F)BiS$_2$ single crystals was confirmed as an intrinsic property, which has previously only been reported for polycrystalline samples [3,26].

## 2. Experimental Procedures

### 2-1. Growth method for BiS$_2$-layered superconducting single crystals

Single crystals of R(O,F)BiS$_2$ (R = La, Ce, Pr, Nd) were grown by a high-temperature flux method in a vacuumed quartz tube. Alkali metal chlorides were chosen as fluxes, of which CsCl, RbCl, and KCl had the low reactivity with the quartz tube. The melting temperature of CsCl (645 °C) is the lowest of these alkali metal chlorides. Additionally, a CsCl/KCl mixture with a eutectic temperature of approximately 616 °C is reported at the molar ratio of CsCl:KCl = 5:3 [27].

The CsCl/KCl mixture (CsCl:KCl = 5:3) was employed for the flux. The raw materials for the growth of R(O,F)BiS$_2$ single crystals were R$_2$S$_3$, Bi, Bi$_2$S$_3$, Bi$_2$O$_3$, BiF$_3$, CsCl, and the CsCl/KCl mixture (CsCl:KCl = 5:3) flux [27]. The starting materials were weighed to obtain a nominal composition of RO$_{1-x}$F$_x$BiS$_2$ ($x$ = 0-1.0). A mixture of raw materials (0.8 g) and the CsCl/KCl flux (5.0 g) were combined using a mortar, and then sealed in a quartz tube under vacuum (~10 Pa). This mixed powder was heated at 800 °C for 10 h, cooled slowly to 600 °C at a rate of 1 °C/h, and then furnace-cooled to room temperature (around 20-30 °C), except for LaO$_{0.9}$F$_{0.1}$BiS$_2$ ($x$ = 0.1) and the F-free

LaOBiS$_2$, CeOBiS$_2$ and PrOBiS$_2$ single crystals. For LaO$_{0.9}$F$_{0.1}$BiS$_2$, LaOBiS$_2$ and PrOBiS$_2$, the maximum heat treatment temperature was 1000 °C; other conditions were maintained. In PrOBiS$_2$, the obtained single crystals were not only PrOBiS$_2$ but also Bi$_2$S$_3$. Meanwhile, the flux for the growth of the CeOBiS$_2$ single crystal was only 5.0 g CsCl; the mixed powder was heated at 950 °C for 10 h and cooled slowly to 650 °C at a rate of 1 °C/h. The quartz tube was opened in air, and the flux was dissolved in the quartz tube using distilled water. The remaining product was then filtered and washed with distilled water. R(O,F)BiS$_2$ single crystals were collected from the product.

**2-2. Characterization of obtained single crystals**

  The crystal structure, morphology, and composition of the obtained single crystals were evaluated by X-ray diffraction (XRD) using CuK$_\alpha$ radiation, scanning electron microscopy (SEM), and electron probe microanalysis (EPMA). Single-crystal XRD data were collected using a Rigaku XTALAB-MINI diffractometer with graphite-monochromated MoK$_\alpha$ radiation. The data were corrected for the Lorentz and polarization effects. The crystal structure was solved and refined using computer programs from the Crystal Structure crystallographic software package [28]. The crystal structure was then drawn by VESTA computer software [29]. The superconducting properties of the grown single crystals were also evaluated. The superconducting transition temperature ($T_c$) was estimated from the temperature dependence of magnetization by a superconducting quantum interface device (SQUID) magnetometer with an applied field of 10 Oe. The transport properties of the single crystals were measured by a standard four-probe method with a constant current density ($J$) using a Physical Property Measurement System (Quantum Design; PPMS DynaCool). The electrical terminals were fabricated from Ag (silver) paste. The angular ($\theta$) dependence of resistivity ($\rho$) in the flux liquid state was measured under various magnetic fields ($H$), with the superconducting anisotropy ($\gamma_s$) calculated using the effective mass model [30-32]. For the $c$-axis transport measurement, s-shaped junction of the $Pr(O,F)BiS_2$ single crystal fixed on a single-crystal substrate of $SrTiO_3$ was fabricated by a three-dimensional (3D) focused ion beam (FIB) etching method using a Ga-ion beam [33,34]. Normal-state anisotropy ($\gamma_n$) was estimated from the resistivity-temperature ($\rho$–$T$) characteristics of the as-grown ($J//ab$-plane) and s-shaped junction ($J//c$-axis) samples. The current-voltage ($I$–$V$) characteristics of the s-shaped junction sample were measured by the standard four-probe method. The sample-position (angle between the $ab$-plane and the magnetic field) dependence of the flow voltage ($V_{fl}$) of the s-shaped junction sample was measured under a magnetic field for Josephson vortex research.

The magnetic properties of the Ce(O,F)BiS$_2$ single crystals including the temperature (*T*) dependence of magnetization (*M*) under zero-field cooling (ZFC) and field cooling (FC) were measured by a SQUID magnetometer, with a magnetic field of 10 Oe applied parallel to the *c*-axis. Magnetization-magnetic field (*M–H*) curves with the magnetic field applied parallel to the *c*-axis were also measured.

## 3. Results

### 3.1 Growth of single crystals

R(O,F)BiS$_2$ single crystals appeared with impurities of Bi$_2$S$_3$ and ROCl. The amount of these impurities increased with decreasing the amount of the flux. R(O,F)BiS$_2$ single crystals were found to be plate-like in shape. Figure 1 shows a typical SEM image of the obtained Nd(O,F)BiS$_2$ single crystal, which are 1.0-2.0 mm in size and 10-20 μm in thickness [14]. The decrease of the temperature, the slow cooling time, or both produced very small single crystals. For example, the heat treatment condition of 730 °C for 10 h and then slow cooling to 630 °C at 1 °C/h produced single crystals of around 20 μm. Figure 2 shows the XRD pattern of a well-developed plane in a single crystal grown from the starting materials with the nominal composition of NdO$_{0.7}$F$_{0.3}$BiS$_2$. The XRD pattern of figure 2 shows the presence of only 00$l$ diffraction peaks of the NdOBiS$_2$ crystal structure, indicating that the $ab$-plane is well developed. Thus, the growth rate in the $c$-axis direction is less than that of the $a$-axis direction. The in-plane orientation of the $ab$-plane in the grown crystals was not measured.

  Table I summarizes the nominal F composition ($x$) and the rare earth elements (R) in the starting materials, as well as the analytical F composition ($y$), $c$-axis lattice parameter ($c$), $T_c$, and superconducting anisotropy ($\gamma_s$) in the grown single crystals.

Details of $\gamma_s$ are described in section 3.3. The F composition is normalized by the total F

and O content. The $y$ values in the Ce(O,F)BiS$_2$ single crystals measured by EPMA

were overestimated because the characteristic X-ray signals of F-K$_\alpha$ (677 eV) and

Ce-M$_\zeta$ (676 eV) overlapped [35]. The overlapped signals were measured from the F-free

CeOBiS$_2$ single crystal, which was estimated to be $y = 0.37$. Thus, the $y$ value in the

Ce(O,F)BiS$_2$ single crystals was calculated by subtracting the overlapped signals

estimated from the F-free CeOBiS$_2$ single crystal. Meanwhile, the chemical ratio of

R:Bi:S in the crystals as determined by EPMA was $1 \pm 0.08 : 1 \pm 0.06 : 2$; this ratio is in agreement with stoichiometry. No Cs, K, or Cl was detected in the crystals by EPMA at the minimum sensitivity limit of 0.1 wt%.

  On increasing the value of $x$, the value of $y$ also increases and reaches saturation at

0.45 (R = La), 0.29 (R = Ce), 0.26 (R = Pr), and 0.42 (R = Nd) when $x$ exceeds around

0.6. The increase in $x$ decreases the $c$-axis lattice parameters ($c$), but provides almost equivalent values as when $x \geq 0.6$ (R = Pr, $x \geq 0.5$). The minimum $c$-axis lattice parameter is about 13.4 Å regardless of the R used.

Single crystals of La(O,F)BiS$_2$ and Ce(O,F)BiS$_2$ with $x \leq 0.3$ did not exhibit

superconductivity down to 2 K. When $x \geq 0.4$, superconducting transition temperatures

($T_c$) were observed at around 3 K. Pr(O,F)BiS$_2$ single crystals showed a superconducting transition of about 4 K, except those with $x \leq 0.1$, which could not be measured in this system because of the small size of the obtained crystals. The maximum $T_c$ of approximately 5 K was observed in the Nd(O,F)BiS$_2$ single crystals.

**3.2 Single-crystal X-ray diffraction analysis**

Single-crystal structural analysis of R(O,F)BiS$_2$ was performed using Nd(O,F)BiS$_2$

grown under conditions of $x = 0.3$ and $y = 0.29$. The structural analysis showed that the

crystal structure belonged to the tetragonal space group *P4/nmm* (#129) with $a =$

3.996(3) Å and $c = 13.464(6)$ Å [14]. Structural refinement was performed on the

NdOBiS$_2$ model because of the difficulty in distinguishing between O and F. The final refinement was performed with fixed occupancies of unity for Nd, Bi, S, and O. As shown in figure 3 [14], the crystal structure of $NdOBiS_2$ is composed of stacked $Nd_2O_2$ layers and $Bi_2S_4$ layers, isostructural with $LaOBiS_2$ [2]. The F composition dependence of the crystal structure was revealed by the $La(O,F)BiS_2$ single crystals, in which Bi-S planes became flatter with increasing F contents [36]. The single-crystal structural analysis of $Ce(O,F)BiS_2$ was systematically performed, which provided more details on the relationship between F composition and crystal structure [37]. The structural parameters of single crystals the $Ce(O,F)BiS_2$ grown under conditions of $x = 0.3$-$0.9$ and $y = 0.16$-$0.29$ are summarized in table II [37]. The definition of atomic positions in $Ce(O,F)BiS_2$ is shown in figure 4 [37]. According to table II, while the increase in F content does not significantly affect the Bi−S(1) bond length, it does change the S(1)−Bi−S(1) angle. In samples with high F concentrations (Ex: Crystal 3-5), this angle is nearly flat, whereas in a low-F concentration sample (Ex: Crystal 1) it is slightly zigzag. These changes are consistent with the results of the single-crystal analysis of $La(O,F)BiS_2$ [36]. The increase in F concentration increases the Bi−S(2) and Ce−(O,F) bond distances, but decreases the Ce−S(2) bond distance. This agrees with the reported extended X-ray absorption fine structure (EXAFS) analysis of powder samples of $Ce(O,F)BiS_2$ [38]. The angles of the S(1)−Bi−S(2) and Ce−(O,F)−Ce bonds decreased with increasing F concentration. On the other hand, the S(2)−Ce−S(2) bond angle increased.

### 3.3 Superconducting anisotropies of BiS$_2$-layered single crystals

The temperature dependence of resistivity in the R(O,F)BiS$_2$ single crystals differed significantly under variation in the angle of an applied magnetic field $H$, between parallel to the $ab$-plane and to the $c$-axis. This suggests that R(O,F)BiS$_2$ is a high-anisotropy material. Figure 5 shows the temperature dependence of resistivity in a single crystal of Nd(O,F)BiS$_2$ grown under the conditions of $x = 0.3$ and $y = 0.29$ at temperatures below 10 K under $H = 0$-9.0 T, parallel to the (a) $ab$-plane and (b) $c$-axis [14]. The suppression of the critical temperature under $H$ applied parallel to the $c$-axis is more significant than that under $H$ parallel to the $ab$-plane. The field dependences of $T_c^{onset}$ and $T_c^{zero}$ under $H$ parallel to the $ab$-plane ($H$//$ab$-plane) and to the $c$-axis ($H$//$c$-axis) are plotted in figure 6 [14]. The linear extrapolations of $T_c^{onset}$ for the cases of $H$//$ab$-plane and $H$//$c$-axis approach the values of 42 and 1.3 T, respectively. Therefore, the upper critical fields $H^{//ab}_{C2}$ and $H^{//c}_{C2}$ at zero temperature are estimated to be 29 and 0.90 T, as determined by the Werthamer-Helfand-Hohenberg (WHH) theory [44]:

$$H_{C2}(0) = -0.693T_c(dH_{C2}/dT)_{T_c} \qquad (1)$$

In a BCS superconductor in the weak-coupling limit, the Pauli limit is $H_p = 1.84T_c$ [45], which corresponds to 9.4 T ($T_c = 5.1$ K) for the single-crystal Nd(O,F)BiS$_2$ grown under conditions of $x = 0.3$ and $y = 0.29$. In conventional superconductor, $H_{C2}(0)$ is limited by Pauli limit as the $H_p$. However, the value of $H^{//ab}_{C2}$ was 29 which is larger than the $H_p$. This result indicates R(O,F)BiS$_2$ superconductors are possibility of unconventional superconductor. By linear fitting the $T_c{}^{zero}$ data, the irreversibility fields $H^{//ab}_{irr}$ and $H^{//c}_{irr}$ are found to be 16 and 0.64 T, respectively. This indicates that the Nd(O,F)BiS$_2$ superconductor has high anisotropy. The superconducting anisotropy ($\gamma_s$) as evaluated from the ratio of the upper critical field, using the equation:

$$\gamma_s = H^{//ab}_{C2}/H^{//c}_{C2} = \xi_{ab}/\xi_c \quad (\xi: \text{coherence length}) \tag{2}$$

was 32.2 for the single-crystal Nd(O,F)BiS$_2$ grown under conditions of $x = 0.3$ and $y = 0.29$.

Moreover, the superconducting anisotropy of superconducting Nd(O,F)BiS$_2$ was estimated using the effective mass model. The angular ($\theta$) dependence of resistivity ($\rho$) was measured under different $H$ in the flux liquid state to estimate $\gamma_s$, as reported in Refs. 30 and 31. The reduced field ($H_{red}$) was calculated using the following equation for an effective mass model:

$$H_{red} = H(\sin^2\theta + \gamma_s^{-2}\cos^2\theta)^{1/2} \tag{3}$$

where $\theta$ is the angle between the $ab$-plane and the magnetic field [32] and $H_{red}$ is calculated from $H$ and $\theta$. The superconducting anisotropy ($\gamma_s$) was estimated from a best scaling of the $\rho$–$H_{red}$ relationship. Figure 7 displays the $\theta$ dependence of $\rho$ for $H =$ 0.1-9.0 T in the flux liquid state for a single crystal of Nd(O,F)BiS$_2$ grown under conditions of $x = 0.3$ and $y = 0.29$ [14]. Small dips are observed in the $\rho$–$\theta$ curves around the $H$//$c$-axis at $H < 0.6$ T. These dips presumably originate from small cracks, which behave as weak pinning sites in the single crystal. The $\rho$–$\theta$ curve exhibits a twofold symmetry. Figure 8 shows the $\rho$–$H_{red}$ scaling obtained from the $\rho$–$\theta$ curves in figure 7 using Eq. (3) [14]. The scaling was performed by taking $\gamma_s = 30$, as shown in figure 8. The value of $\gamma_s$ estimated from the effective mass model was consistent with that of using Eq. (2). The $\gamma_s$ of R(O,F)BiS$_2$ single crystals with various $x$ and R elements were estimated from the effective mass model, as summarized in table I. The $\gamma_s$ of the crystal was estimated to be 30–60 when R = La, Pr, and Nd; the value of $\gamma_s$ was especially high for R = Pr, while it was lower than expected (13–21) for R = Ce. The origin of this variation is unclear.

### 3.4 *C*-axis transport properties of PrO$_{1-x}$F$_x$BiS$_2$ single crystals

Single crystals are suitable for anisotropic measurements. The differences in transport properties between in-plane (*ab*-plane) and out-of-plane (*c*-axis) using the Pr(O,F)BiS$_2$ single crystal are demonstrated in this chapter. The superconducting anisotropies of single crystals of Pr(O,F)BiS$_2$ grown under conditions of $x$ = 0.3 and $y$ = 0.18 were estimated to be around 55, demonstrating the potential use of the crystals as intrinsic Josephson junctions along the *c*-axis [24,25]. Intrinsic Josephson junctions emerge in the crystal structures of layered superconductors. The s-shaped junction of the Pr(O,F)BiS$_2$ single crystal was fabricated for the *c*-axis transport measurement [33,34]. Figure 9 shows a scanning ion microscopy (SIM) image along the *c*-axis of the s-shaped junctions fabricated on the Pr(O,F)BiS$_2$ single crystal with $x$ = 0.3 and $y$ = 0.18 [39]. The cross-sectional area and thickness of the junctions are 5.4 × 5.7 μm$^2$ along the *ab*-plane and 1.0 μm, respectively. The direction of current flow is shown in figure 9, along the *c*-axis in the junction. This junction is used to characterize the transport properties along the *c*-axis. Figure 10 shows the $\rho$–$T$ characteristics along the (a) *ab*-plane ($\rho_{ab}$) and (b) *c*-axis ($\rho_c$) of the Pr(O,F)BiS$_2$ single crystal with $x$ = 0.3 and $y$ = 0.18, as measured by the standard four-probe method [39]. The normal-state anisotropies $\gamma_n$ = $(\rho_c/\rho_{ab})^{1/2}$ at 5 and 250 K are found to be 39.3 and 21.1, respectively. These $\gamma_n$ values are lower than the $\gamma_s$ values.

Figure 11 shows the *I*–*V* characteristics along the *c*-axis (shown in figure 9) at 2.0 K

[39]. The critical current density ($J_c$) is approximately $1.33 \times 10^3$ A/cm$^2$ in the self-field. The $I$–$V$ curve shows a hysteresis. However, the multi-branched structure of the $I$–$V$ characteristics corresponding to intrinsic Josephson junctions [33,34] is not observed in the $I$–$V$ curve of the s-shaped junction of the Pr(O,F)BiS$_2$ single crystal with $x = 0.3$ and $y = 0.18$.

The sample-position (the angle between the $ab$-plane and the magnetic field) dependence of the $V_{ff}$ characteristics under $H$ for s-shaped single-crystal junction (Figure 9) was measured. Flow resistance was defined as $V_{ff}$ divided by the direct-current (dc) bias current ($I$) to conduct the measurements. Figure 12 shows the sample-position dependence of the $V_{ff}$ characteristics at various $I$ values along the $c$-axis of the s-shaped junctions under 1.0 T magnetic field at 2.0 K [39]. Figure 12(b) shows that $V_{ff}$ reaches a local maximum under a magnetic field parallel to the $ab$-plane. The local maximum $V_{ff}$ increases with increasing applied current. An increase in the local maximum $V_{ff}$ is also observed in the $I$–$V$ characteristics under $H$ parallel to the $ab$-plane, as shown in figure 12(c). This phenomenon can be explained by the "lock-in" state [40,41], which may originate from the Josephson vortex flow. This result indicates that pancake vortices appear in the Pr(O,F)BiS$_2$ single crystal with $x = 0.3$ and $y = 0.18$ due to the high superconducting anisotropy of the crystal. $V_{ff}$ markedly decreases under $H$ approximately parallel to the *ab*-plane, suggesting the dissipation of these pancake vortices. Subsequently, Josephson vortices appear under $H$ parallel to the *ab*-plane, indicating the "lock-in" state. The lock-in state is expected to be free from pancake vortices, which cross the superconducting $BiS_2$ layers.

**3.5 Magnetic properties of $CeO_{1-x}F_xBiS_2$ single crystals**

Ce(O,F)BiS$_2$ exhibits the coexistence of superconductivity and magnetism in polycrystalline samples [3,26]. Although the existence of both superconductivity and magnetic ordering in mixed anion compounds has been observed by different groups, the values reported for these properties have differed. This may be attributed to the inhomogeneity of powder samples or the effects of impurity phases and grain boundaries on the sample characterization. As only powder samples have been analyzed, it is important to further examine the structural and property details in single-crystal analysis [37]. Figure 13 shows the temperature dependence of the resistivity ($\rho$–$T$) and magnetization ($M$–$T$) for single crystals of Ce(O,F)BiS$_2$ grown under conditions of $x$ = 0.3-0.9 and $y$ = 0.16-0.29, between 2−10 K [37]. In the $\rho$–$T$ characteristics, all crystals exhibit semi-conductive behavior in the normal region, consistent with previous reports of Ce(O,F)BiS$_2$ powders with different F concentrations [3,6,26].

The $M-T$ curve of Ce(O,F)BiS$_2$ single crystals with $x = 0.3$ and $y = 0.16$ shows no clear transition. The single crystals of Ce(O,F)BiS$_2$ grown under conditions of $x = 0.5$-0.9 and $y = 0.24$-0.29 show a decrease in magnetization at around 3 K, which corresponds to the superconducting transition temperature ($T_c$). Additionally, a deviation between ZFC and FC was found in Ce(O,F)BiS$_2$ single crystals with $x = 0.5$ and $y = 0.24$ near 5 K, which suggests spontaneous magnetization. This deviation is enhanced, and its onset shifted to higher temperatures, with increasing F concentration; in Ce(O,F)BiS$_2$ single crystals with $x = 0.9$ and $y = 0.29$, the onset is near 7 K. These temperatures of magnetic ordering were defined as $T_m$, which are summarized in table II. Further characterization of these single crystals is required in order to investigate the origin of this phenomenon.

Figure 14 shows the $M-H$ curve of the Ce(O,F)BiS$_2$ single crystals with $x = 0.9$ and $y = 0.29$ at 2 K [37]. The hysteresis loop observed supports the theory of spontaneous magnetization. The saturating magnetization was ca. 0.02 μB/Ce [37], much smaller than that expected for the Ce$^{3+}$ free ion (2.54 μB/Ce). The value of the lower critical field ($H_{c1}$) for superconductivity, attributed to the superconductivity, is 10−20 Oe.

## 4. Conclusion

The growth of R(O,F)BiS$_2$ (R = La, Ce, Pr, Nd) single crystals was achieved by using CsCl/KCl flux containing varying rare earth metals and F compositions. The characterization of the obtained single crystals clarified the detailed crystal structures, transport properties, and magnetic properties. An increase in the nominal F compositions in the starting materials was found to enhance the values of F concentration in the obtained single crystals. On increasing the nominal F composition, the analytical F composition also increased and reached saturation at 0.45 (R = La), 0.29 (R = Ce), 0.26 (R = Pr), and 0.42 (R = Nd) when the nominal F composition exceeded around 0.6. The Bi-S planes of the R(O,F)BiS$_2$ single crystals became flatter with increasing F concentration, according to the single-crystal X-ray diffraction analysis. The superconducting anisotropies of the obtained single crystals were estimated to be 30–60 when R = La, Pr and Nd. The anisotropy was especially high for R = Pr, whereas the value was lower than expected (13–21) when R = Ce. Ce(O,F)BiS$_2$ single crystals exhibited a magnetic order at about 5-7 K in addition to the superconducting transition at approximately 3 K, which suggested that the coexistence of superconductivity and magnetism in Ce(O,F)BiS$_2$ was intrinsic. The s-shaped junction of Pr(O,F)BiS$_2$ single crystal with $x = 0.3$ and $y = 0.18$ under a magnetic field parallel to the $ab$-plane of the crystal exhibited a "lock-in" state for Josephson vortex flow.

**Acknowledgments**

This work was supported by JSPS KAKENHI (Grant-in-Aid for challenging Exploratory Research) Grant Number 15K14113.

The author would like to thank Dr. A. Miura (Hokkaido University) for useful discussion and critical reading, Prof. S. Watauchi, Prof. I. Tanaka, Prof. T. Takei, Prof. N. Kumada (University of Yamanashi), Dr. Y. Takano, Dr. M. Tanaka (National Institute for Materials Science), Dr. Y. Mizuguchi (Tokyo Metropolitan University), Dr. S. Demura (Tokyo University of Science), Dr. K. Deguchi (JECC Torisha Co., Ltd.), Dr. M. Fujioka (Hokkaido University), and Dr. H. Okazaki (Okayama University) for their valuable advice.

**Figure Legends and Table Captions**

Figure 1. Typical SEM image of $Nd(O,F)BiS_2$ single crystal [14].

Figure 2. XRD pattern of well-developed plane of $Nd(O,F)BiS_2$ single crystal grown from the starting material with nominal composition of $NdO_{0.7}F_{0.3}BiS_2$.

Figure 3. (Color) Crystal structure of $NdOBiS_2$. The dashed line indicates the unit cell [14].

Figure 4. (Color) Crystal structure of $Ce(O,F)BiS_2$ [37]. "Reprinted with permission from Crystal Growth Design 2015, vol.15, pp.39−44. Copyright 2015 American Chemical Society."

Figure 5. (Color) Temperature dependence of resistivity for a single crystal $Nd(O,F)BiS_2$ grown under conditions of $x = 0.3$ and $y = 0.29$ at magnetic fields of 0-9.0 T parallel to the (a) $ab$-plane and (b) $c$-axis [14].

Figure 6. Data in figure 5 after plotting of field dependences of $T_c^{onset}$ and $T_c^{zero}$ under magnetic fields ($H$) parallel to the $ab$-plane ($H$//$ab$-plane) and $c$-axis ($H$//$c$-axis). The lines are linear fits to the data. The inset is an enlargement of the lower-field region [14].

Figure 7. (Color) Angular $\theta$ dependence of resistivity $\rho$ in flux liquid state at various

magnetic fields (bottom to top, 0.1 to 9.0 T) for a single crystal $Nd(O,F)BiS_2$ grown under conditions of $x = 0.3$ and $y = 0.29$ [14].

Figure 8. (Color) Data in figure 7 after scaling of angular $\theta$ dependence of resistivity $\rho$ at a reduced magnetic field of $H_{red} = H(\sin^2\theta + \gamma_s^{-2}\cos^2\theta)^{1/2}$ [14].

Figure 9. SIM image of s-shaped junction. The cross-sectional area and thickness of the junctions are about 30.78 $\mu m^2$ and 1.0 $\mu m$, respectively [39]. "Copyright 2015 by The Japan Society of Applied Physics"

Figure 10. $\rho$–$T$ characteristics of $Pr(O,F)BiS_2$ single crystal with $x = 0.3$ and $y = 0.18$ along the (a) $ab$-plane ($\rho_{ab}$) and (b) $c$-axis ($\rho_c$). Sample for $c$-axis transport measurement is shown in figure 9 [39]. "Copyright 2015 by The Japan Society of Applied Physics"

Figure 11. $I$–$V$ characteristics of s-shaped $Pr(O,F)BiS_2$ single crystal with $x = 0.3$ and $y = 0.18$ junction at 2.0 K and self-field. $J_c$ is about $1.33 \times 10^3$ $A/cm^2$ [39]. "Copyright 2015 by The Japan Society of Applied Physics"

Figure 12. (a) Sample-position (angle between $ab$-plane and magnetic field) dependence of flow voltage $V_{ff}$ at various currents ($I$) for s-shaped $Pr(O,F)BiS_2$ single crystal with $x = 0.3$ and $y = 0.18$ junction under 1.0 T magnetic fields at 2.0 K. (b) Enlargement of sample position approximately parallel to $ab$-plane. (c) $I$–$V$ characteristics of s-shaped junction at 2.0 K and 1.0 T magnetic field parallel to $ab$-plane [39]. "Copyright 2015 by

Figure 13. (Color) Resistivity-temperature characteristics (top) and magnetization (bottom) of single crystals $Ce(O,F)BiS_2$ grown under conditions of $x = 0.3$-$0.9$ and $y = 0.16$-$0.29$. The resistivity values have been normalized to those acquired at 10 K. The temperature dependence of the magnetization was measured under zero-field cooling (ZFC) and field cooling (FC) with an applied magnetic field of 10 Oe parallel to the $c$-axis [37]. "Reprinted with permission from Crystal Growth Design 2015, vol.15, pp.39−44. Copyright 2015 American Chemical Society."

Figure 14. $M−H$ curve of $Ce(O,F)BiS_2$ single crystals with $x = 0.9$ and $y = 0.29$ at 2 K under an applied magnetic field parallel to the $c$-axis [37]. "Reprinted with permission from Crystal Growth Design 2015, vol.15, pp.39−44. Copyright 2015 American Chemical Society."

Table I   Dependence of nominal F composition in the starting materials ($x$) on the analytical F composition ($y$), $c$-axis lattice parameter ($c$), superconducting transition temperature ($T_c$) and superconducting anisotropy ($\gamma_s$) in the grown single crystals.

| | | Nominal F compositions ($x$) in $RO_{1-x}F_xBiS_2$ | | | | | | | | | |
|---|---|---|---|---|---|---|---|---|---|---|---|
| | | 0 | 0.1 | 0.2 | 0.3 | 0.4 | 0.5 | 0.6 | 0.7 | 0.8 | 0.9 |
| R=La | $y$ | 0* | * | 0.24 | 0.23 | 0.37 | 0.43 | 0.45 | 0.46 | 0.44 | 0.46 |
| | $c$ (Å) | 13.86* | 13.63* | 13.58 | 13.57 | 13.44 | 13.42 | 13.39 | 13.37 | 13.37 | 13.39 |
| | $T_c$ (K) | — | — | — | — | 2.7 | 3.3 | 3.1 | 3.0 | 3.3 | 3.2 |
| | $\gamma_s$ | — | — | — | — | 34 | 32-34 | 29-35 | 23-37 | 36 | 36-45 |
| R=Ce | $y$ | 0* | 0.05[#] | 0.16[#] | 0.16[#] | 0.20[#] | 0.24[#] | 0.26[#] | 0.28[#] | 0.27[#] | 0.29[#] |
| | $c$ (Å) | 13.59* | 13.57 | 13.55 | 13.54 | 13.48 | 13.45 | 13.40 | 13.39 | 13.38 | 13.40 |
| | $T_c$ (K) | — | — | — | — | 2.1 | 3.0 | 3.2 | 3.1 | 3.1 | 2.9 |
| | $\gamma_s$ | — | — | — | — | 17 | 15-21 | 13-16 | 13-21 | 11-12 | 14 |
| R=Pr | $y$ | 0* | 0.05 | 0.13 | 0.18 | 0.23 | 0.26 | 0.26 | 0.26 | 0.28 | --- |
| | $c$ (Å) | 13.80* | 13.67 | 13.55 | 13.49 | 13.41 | 13.39 | 13.38 | 13.37 | 13.37 | --- |
| | $T_c$ (K) | | | 2.4 | 3.8 | 4.0 | 4.1 | 4.3 | 4.4 | 4.6 | --- |
| | $\gamma_s$ | | | 20 | 53-58 | 32-46 | 42-56 | 47 | 38-39 | | --- |
| R=Nd | $y$ | --- | --- | --- | 0.29 | 0.37 | 0.37 | 0.42 | 0.41 | 0.42 | 0.40 |
| | $c$ (Å) | --- | --- | --- | 13.56 | 13.47 | 13.48 | 13.41 | 13.43 | 13.41 | 13.41 |
| | $T_c$ (K) | --- | --- | --- | 5.1 | 4.8 | 5.0 | 5.1 | 5.2 | 5.2 | 5.3 |
| | $\gamma_s$ | --- | --- | --- | 30-34 | 26-32 | 25-30 | 37-40 | 30-31 | 33 | 37-40 |

---:No RO(F)BiS$_2$ single crystals were obtained.    —:Unmeasurable at our system.    *:LaO$_{0.9}$F$_{0.1}$BiS$_2$ ($x$ = 0.1), F-free ($x$ = 0) LaOBiS$_2$,

CeOBiS$_2$ and PrOBiS$_2$ single crystals were own growth conditions. (See to 2.1 chapters.) [#]:The analytical F composition ($y$) in the

Ce(O,F)BiS$_2$ single crystals was subtracted from the overlapped signals of F-free CeOBiS$_2$ single crystal. (See to 3.1 chapters.)

Table II  Summary of F compositions, room-temperature structural parameters, and transition temperatures of superconductivity and magnetic ordering in Ce(O,F)BiS$_2$ single crystals [37]. "Reprinted with permission from Crystal Growth Design 2015, vol.15, pp.39−44. Copyright 2015 American Chemical Society."

| | Ce(O,F)BiS$_2$ | | | | |
|---|---|---|---|---|---|
| Name of crystal | Crystal 1 | Crystal 2 | Crystal 3 | Crystal 4 | Crystal 5 |
| Nominal F composition ($x$) | 0.3 | 0.5 | 0.7 | 0.8 | 0.9 |
| Analytical F composition ($y$) | 0.16 | 0.24 | 0.28 | 0.27 | 0.29 |
| $a$-axis lattice parameters /Å | 4.019(3) | 4.031(3) | 4.026(3) | 4.031(3) | 4.030(3) |
| $c$-axis lattice parameters /Å | 13.507(5) | 13.383(5) | 13.340(5) | 13.338(5) | 13.336(5) |
| Volume /Å$^3$ | 218.2(3) | 217.5(3) | 216.2(3) | 216.7(3) | 216.6(3) |
| Bi–S(1) [interplane] ×4 /Å | 2.843(2) | 2.850(2) | 2.847(2) | 2.850(2) | 2.850(2) |
| S(1)–Bi–S(1)[interplane] /deg | 177.1(4) | 179.3(3) | 180.1(3) | 180.4(4) | 180.4(4) |
| Bi–S(2) /Å | 2.507(8) | 2.517(7) | 2.529(6) | 2.526(8) | 2.537(7) |
| S(1)–Bi–S(2) /deg | 91.4(2) | 90.36(16) | 89.97(16) | 89.8(2) | 89.80(19) |
| Ce–(O,F) ×4 /Å | 2.3826(18) | 2.4044(16) | 2.4048(16) | 2.4098(17) | 2.4112(16) |
| Ce–(O,F)–Ce /deg | 115.00(10) | 113.77(9) | 113.67(9) | 113.52(9) | 113.37(9) |
| Ce–S(2) ×4 /Å | 3.110(4) | 3.096(3) | 3.082(3) | 3.085(4) | 3.079(3) |
| S(2)–Ce–S(2) /deg | 80.50(12) | 81.25(10) | 81.57(9) | 81.57(12) | 81.74(10) |
| S(2)–Ce–(O,F) /deg | 70.93(13) | 70.55(11) | 70.29(9) | 70.34(13) | 70.20(11) |
| Residuals: $R$ ($I > 2.00(I)$) | 0.045 | 0.034 | 0.034 | 0.043 | 0.041 |
| Residuals: $wR2$ (reflections) | 0.091 | 0.075 | 0.084 | 0.084 | 0.077 |
| Goodness of Fit Indicator | 1.03 | 1.03 | 1.16 | 1.08 | 1.19 |
| $T_c$ (superconducting transition) /K | - | 3.0 | 3.1 | 3.1 | 2.9 |
| $T_m$ (magnetic ordering) /K | - | 5.3 | 5.9 | 6.2 | 7.1 |

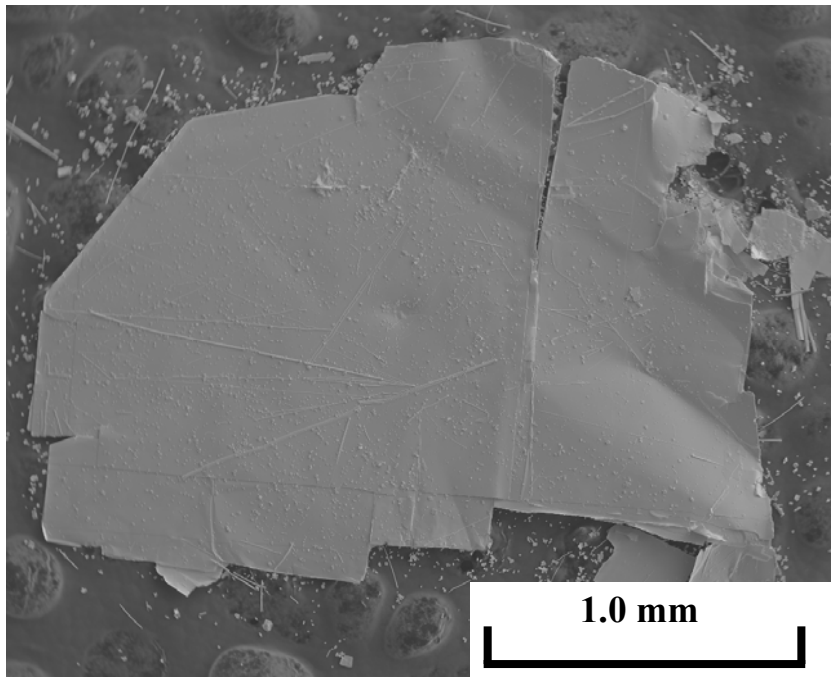

1.0 mm

**Figure 1**

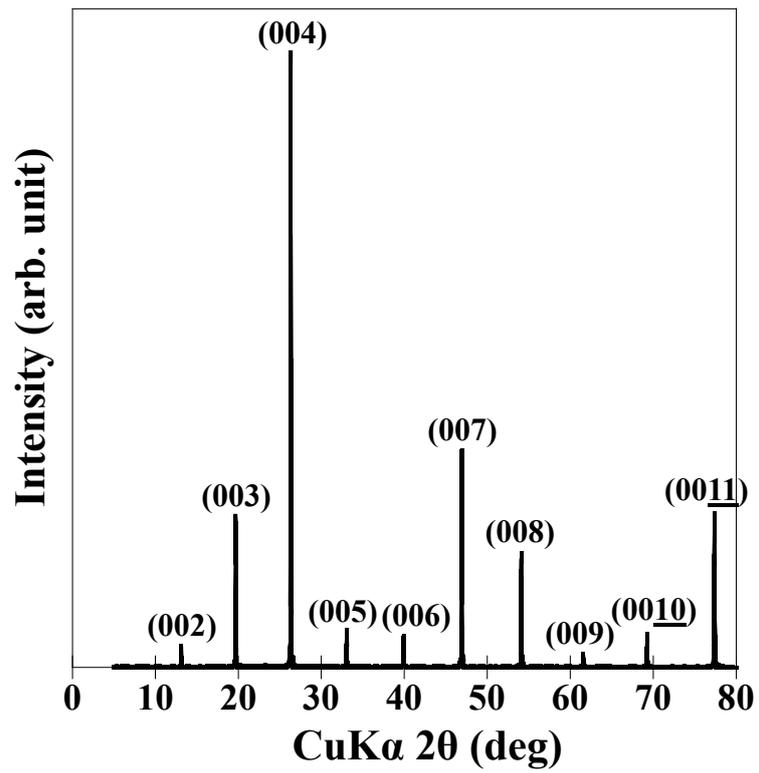

Figure 2

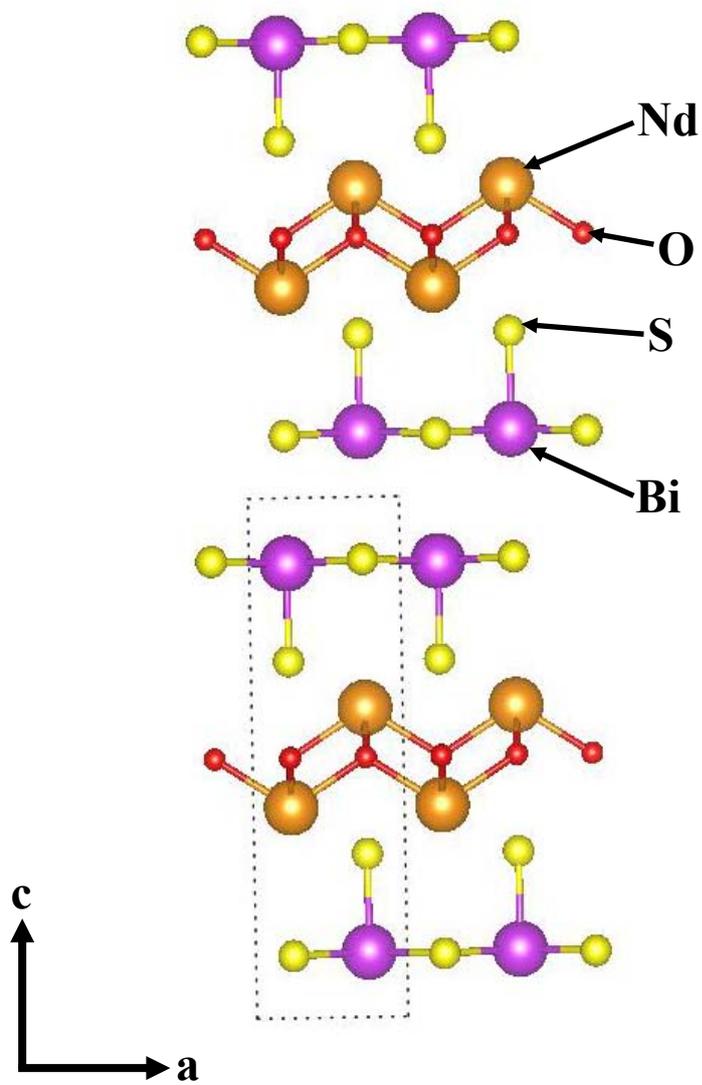

**Figure 3 (Color)**

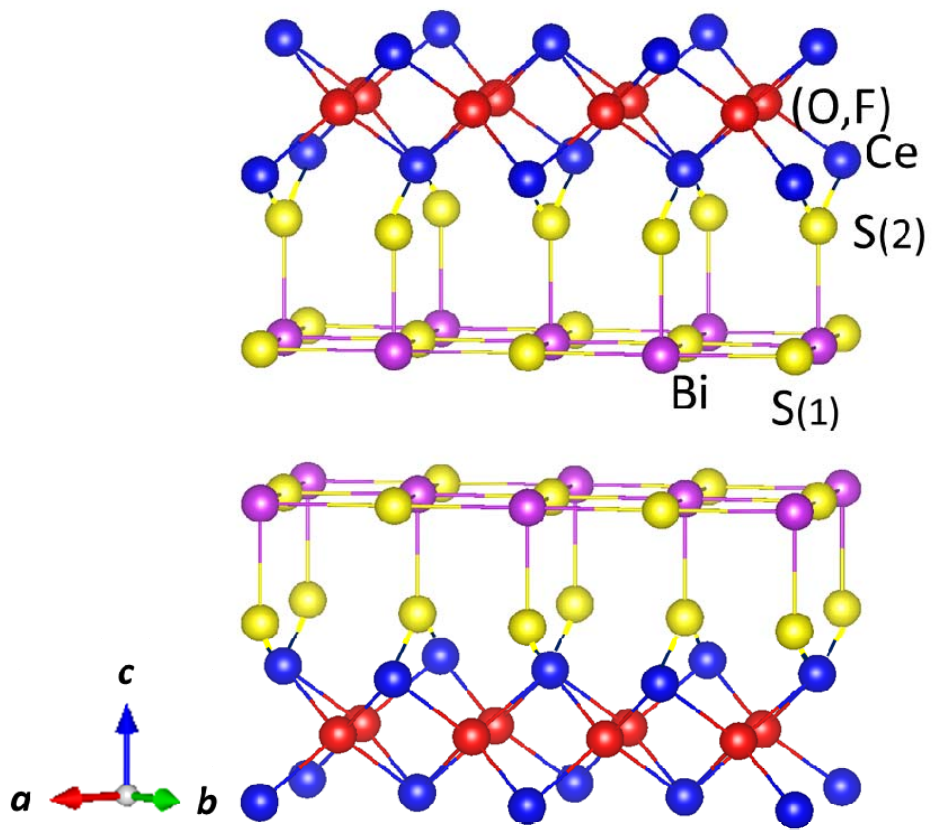

**Figure 4 (Color)**

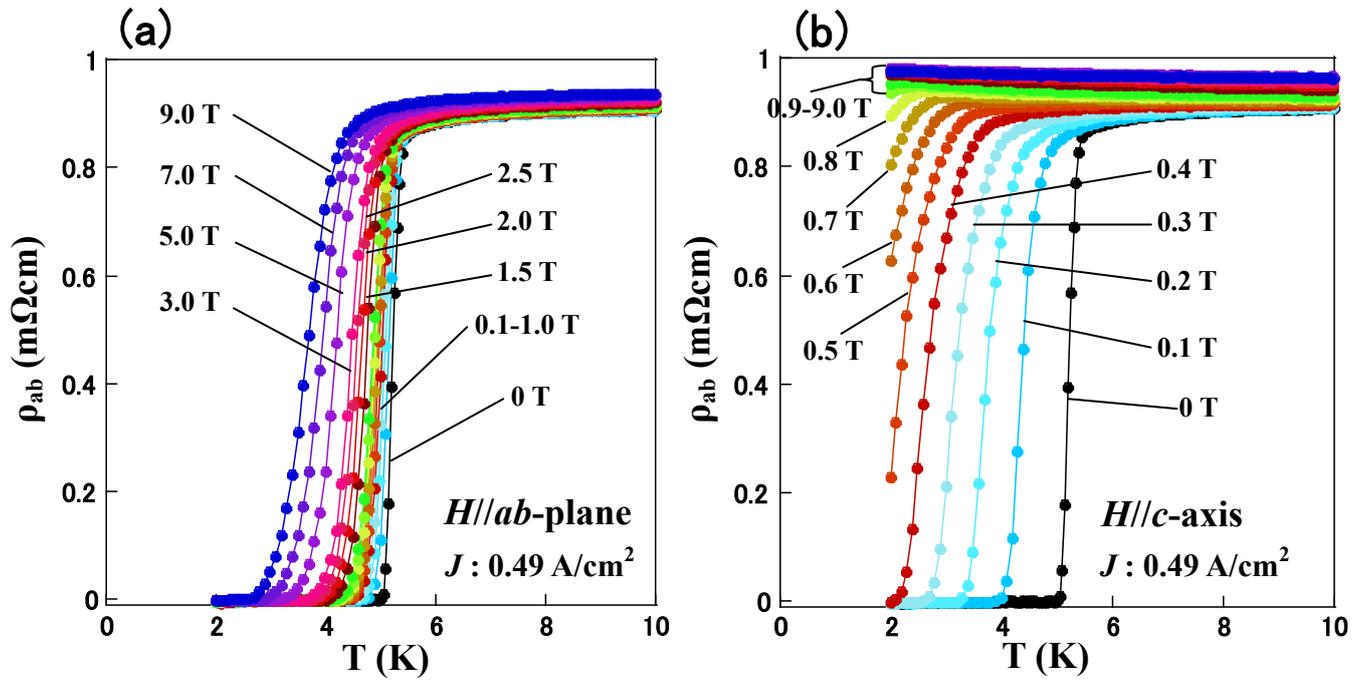

**Figure 5 (Color)**

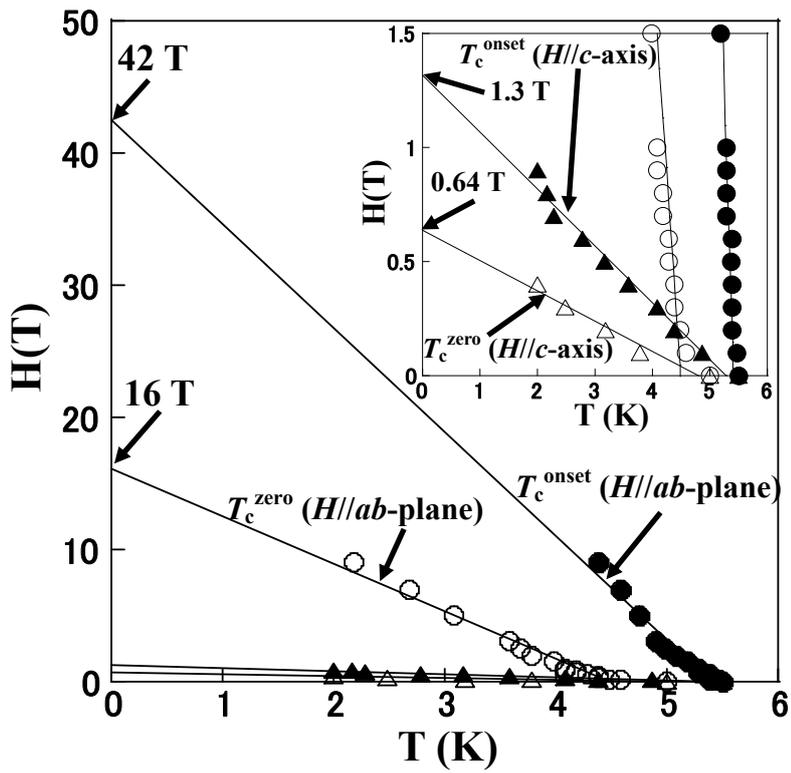

Figure 6

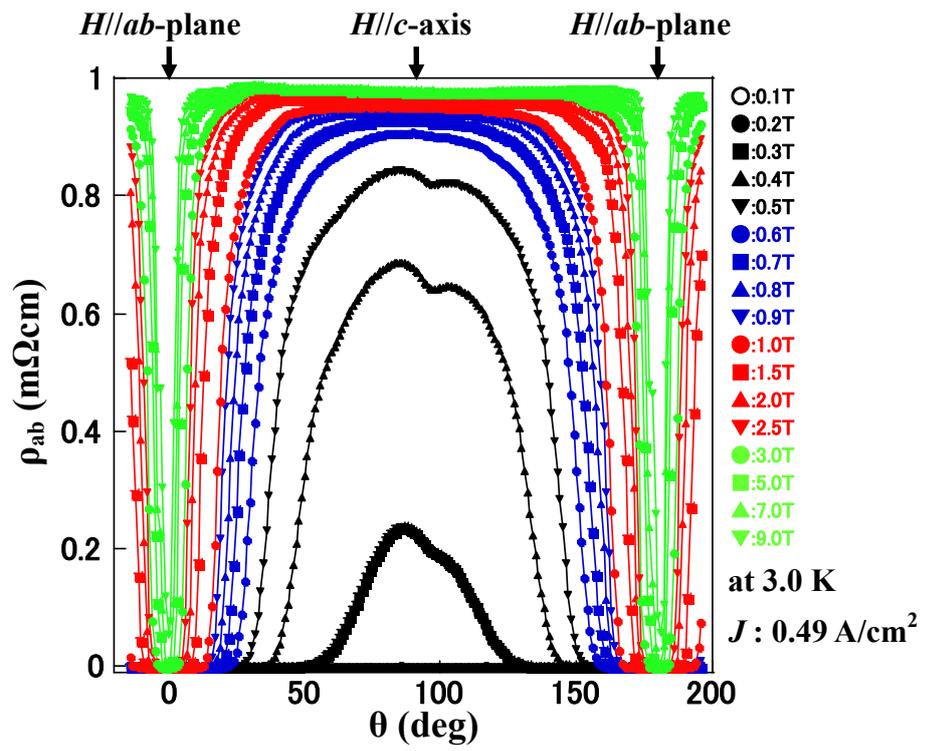

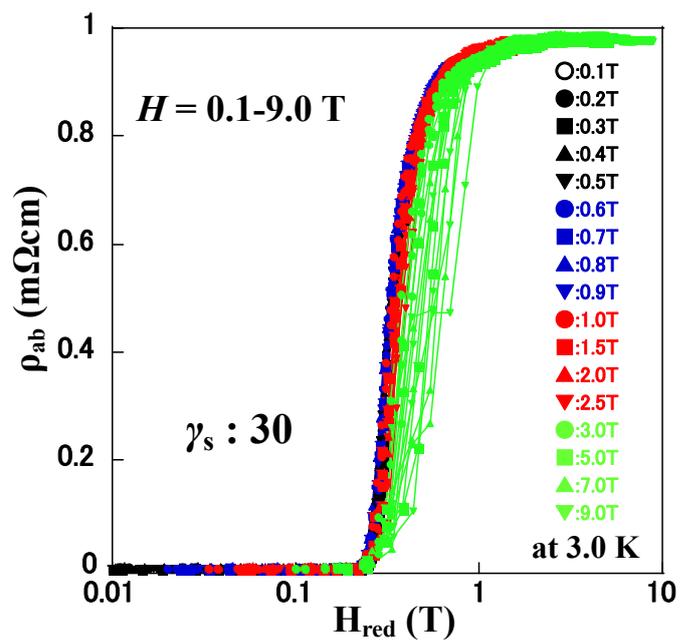

The figure contains the following labels:

$H = 0.1\text{-}9.0$ T

$\rho_{ab}$ (m$\Omega$cm)

$\gamma_s : 30$

- ○:0.1T
- ●:0.2T
- ■:0.3T
- ▲:0.4T
- ▼:0.5T
- ●:0.6T
- ■:0.7T
- ▲:0.8T
- ▼:0.9T
- ●:1.0T
- ■:1.5T
- ▲:2.0T
- ▼:2.5T
- ●:3.0T
- ■:5.0T
- ▲:7.0T
- ▼:9.0T

at 3.0 K

$H_{red}$ (T)

**Figure 8 (Color)**

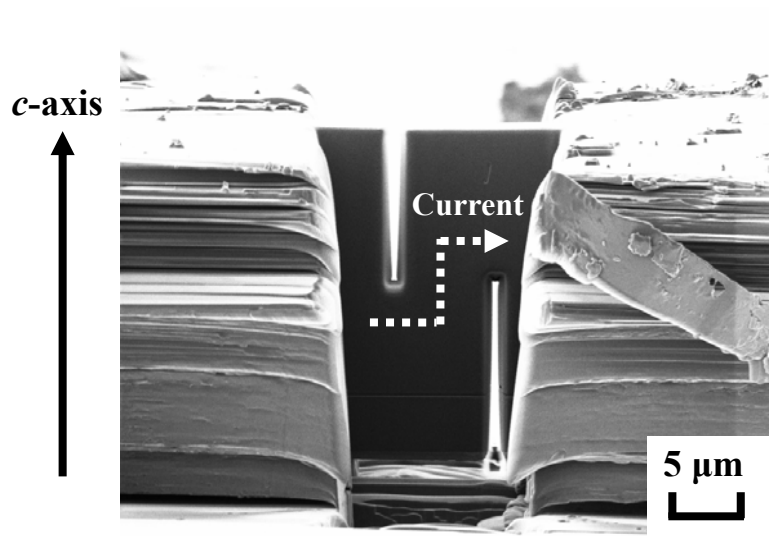

Figure 9

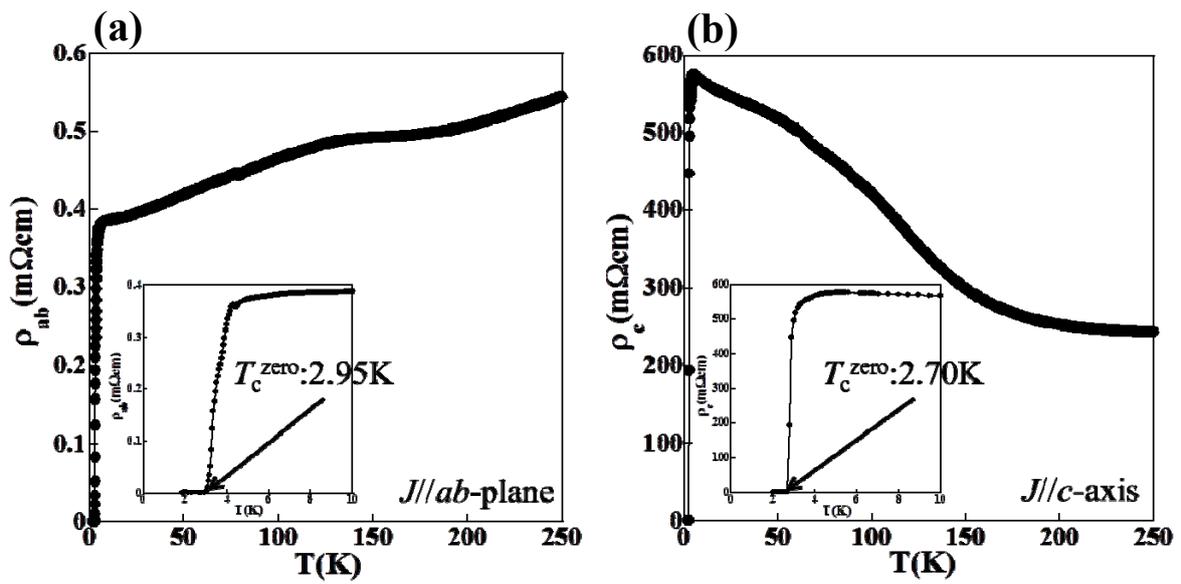

**Figure 10**

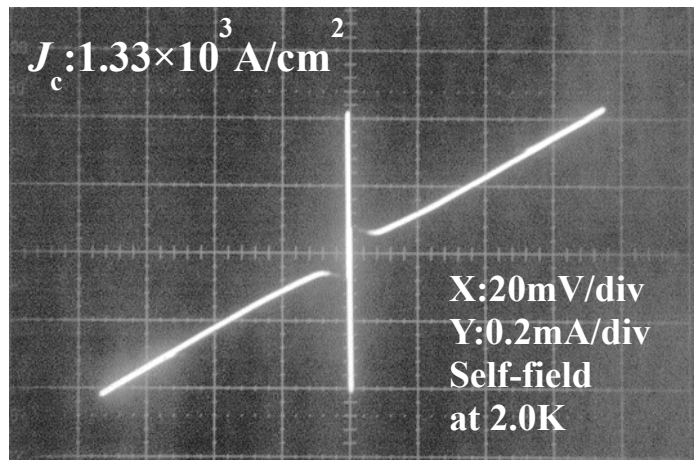

The image contains the following labels:

$J_c$:1.33×10$^3$ A/cm$^2$

X:20mV/div
Y:0.2mA/div
Self-field
at 2.0K

**Figure 11**

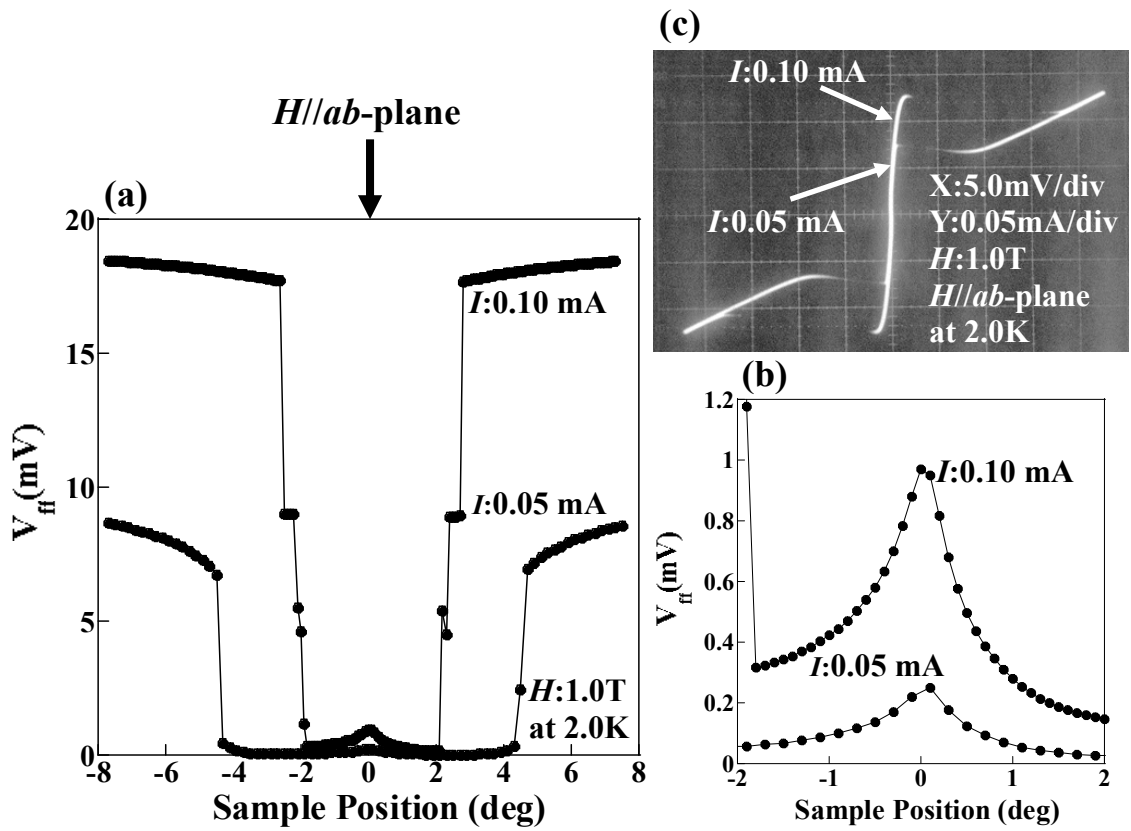

**Figure 12**

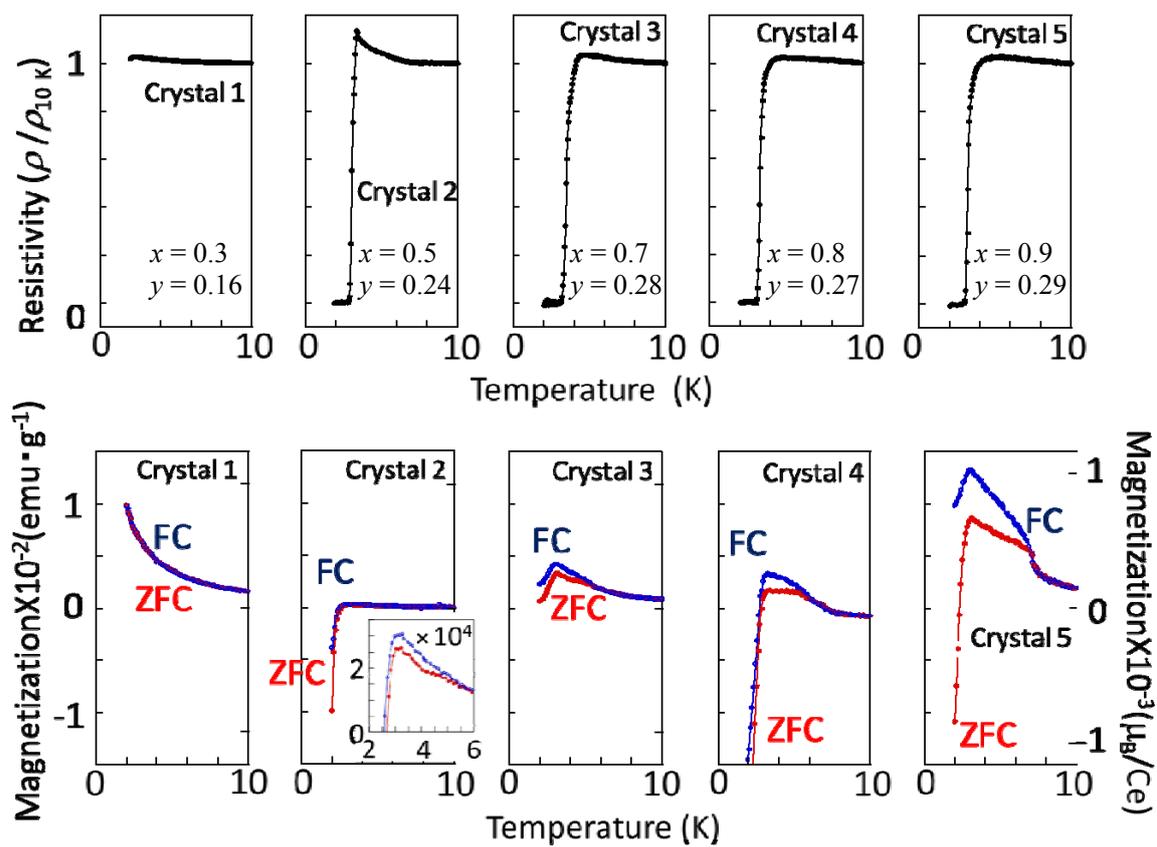

**Figure 13 (Color)**

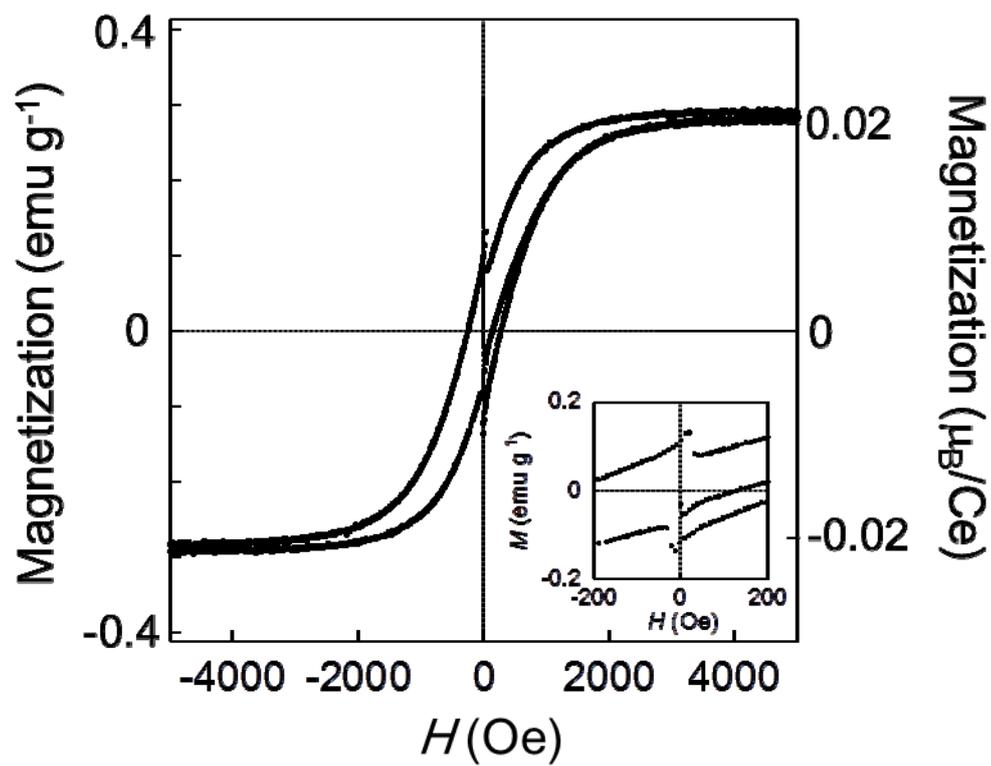

**Figure 14**